# AN ASTRONOMICAL ANALYSIS OF HIPPARCHUS' *COMMENTARIES*

**Gabriele Vanin**
***Independent scholar (rheticus@tiscali.it)***

**Abstract**
Hipparchus is considered the greatest astronomer of antiquity. However, his fame is due, more than to what he has done, or said, or written, to what the scientists of some generation or of a few centuries later have written of him: above all Ptolemy. The obvious discrepancy between the declared period and the real one of the stellar catalogue of the *Almages*t has led to the belief that the catalogue was in reality made by Hipparchus. This factor has contributed greatly to increase his fame, to the detriment of the one of Ptolemy. However, the only work of Hipparchus that we have received, *Commentaries on Aratus and Eudoxus phenomena*, has never been analyzed with the necessary accuracy so far. Such an examination, which is carried out here for the first time, reveals a very different scientist from the one represented by orthodox historiography and is intended as an attempt to avoid the continuation of a singular error, that of judging the value of a scholar not by the existing parts of his work, but only by those that are missing.

## 1 THE FAME OF HIPPARCHUS

For many historians, the figure of Hipparchus stands out as that of the greatest astronomer of antiquity (see for example the classical works of Abetti 1949, and Pannekoek, 1969). He is considered the inspirer or even the author of much of the monumental astronomical work contained in Ptolemy's *Almagest*. This concept originated mainly from the discrepancy, already noted by Arab astronomers, between the age of the star catalogue contained in the *Almagest*, 137 AD, declared by Ptolemy (VII, 4) and the true stellar longitudes reported, referring to 57 AD.[1] Already Tycho Brahe (1610) suggested that the 1° error could be caused by a simple transformation of the coordinates of the lost star catalogue of Hipparchus, of which Pliny speaks (II, 95), using the erroneous precession rate used by Ptolemy, 36" instead 50.3" per year. Brahe's hypothesis was also endorsed by the French astronomers de Lalande and Delambre. The former (1764) assumed that the catalogue of *Almagest* was entirely in Hipparchus, and even wrote that "astronomers are convinced that Ptolemy was only a bad observer, and that all that is good in his work was taken by him from Hipparchus and its predecessors". The latter, who translated the entire catalogue into French in 1817, was the first one to study the question in detail and concluded that many, if not all, the stars of the Ptolemy catalogue were not observed by him. However, in 1796, Pierre-Simon de Laplace found an alternative explanation, which did not involve plagiarism. Over time, many other equally satisfactory explanations were devised (see for example Dobler 2002; Duke 2002, 2002, 2002, 2003; Dreyer 1917, 1918; Evans 1987; Shevchenko 1990; Swerdlow 1992; Vogt 1925) so that modern criticism tends to consider Ptolemy's catalogue as basically original. Nevertheless, it is still widely believed that Hipparchus was a greater astronomer than the great Alexandrian, and that topics and methods of the

[1] The value is the average of the epochs found by Peters and Knobel (1915), 58 AD, and Dobler (2002), 56 AD.

*Almagest* owe very much to the astronomer of Nicaea, and that in some ways Ptolemaic astronomy is a less sophisticated version than the one practiced in the Hellenistic period. Among the recent historians who have been and are of this idea, essentially based on statistical surveys on the star catalogues of Hipparchus and Ptolemy, it is possible to cite, for example, Dambis and Efremov (2000), Graβhof (1990), Newton (1977,1979), Rawlins (1982). Hipparchus' superiority is justified instead for far-reaching considerations, within a framework that offers a very innovative view, even though with controversial points, of the Hellenistic era, in the works of Russo (1994, 1996).

It is remarkable that we do not know almost anything about Hipparchus' life[2] and his activity, if not just as Ptolemy relates us about him, particularly about his observations and his works. Other references to Hipparchus can be found in other authors of ancient times, but, since they are not astronomers, they often have misrepresented or misunderstood his contributions (Neugebauer 1975). Hipparchus wrote a considerable amount of works so that, according to Ptolemy (III, 1), he even compiled their commented catalogue. However, only one of them, *Commentaries on Aratus and Eudoxus phenomena*, arrived to us. According to Toomer (1980) this was due to the fact that, although Hipparchus gained a great reputation in antiquity, his work was not widely read outside the circle of specialists, probably because they were generally in the form of not very long monographs concerning very different subjects and often his works are very technical. Moreover, according to Ptolemy (IX, 2), Hipparchus did not elaborate complete theories or astronomical systems, but rather tested them or elaborated at most some part of them. Thus, when Ptolemy, using part of Hipparchus' work as an essential foundation, built such a system in a book that constituted a true astronomy manual, interest in the original contributions by Hipparchus declined (Toomer 1998). *Commentaries* probably survived thanks to popularity of Aratus' *Phenomena*.

## 2 THE LOST STAR CATALOGUE OF HIPPARCHUS

Regarding this catalogue, the most explicit reference is contained in Pliny (II, 95), who reports that, having Hipparchus noticed a new star, and seeing that he was moving, he began to suspect that the others could do so, and then decided to enumerate and name the stars for posterity by inventing the tools that were suitable for this purpose. This, however, may refer not to a star catalogue, but to a simple list of stars and their positions within each constellation (without any coordinates). As a matter of fact, Ptolemy, wanting to show that the position of the stars had not changed since Hipparchus, reported a list of stellar alignments that presumably Hipparchus had determined (VII, 1) for the purposes cited by Pliny.

In 1892 Ernst Maass published two previously unknown Greek constellation lists, contained in a code of the eighth century, one attributed to Eratosthenes, the other to Hipparchus. They reported nothing but the names of constellations. But in 1898 Alessandro Olivieri (Rehm 1899) and in 1901 Franz Boll discovered two lists that also included the number of stars in each constellation. Later on, other similar lists were discovered, for a total of nine Greeks and two Latins, some anonymous, others attributed to Hipparchus. The most complete is part of Cod. Parisinus 2506, an astrological manuscript of the XIV century, that lists 46 constellations and 653 stars. Since the numbers of stars of only 43 constellations are reported, it can be argued that, if all of them had been reported, as well as stars unlinked between one constellation and the other, the total contained in the original catalogue would have been about 850. Boll and Albert Rehm (Rehm 1899) wrote that it was very probable the Hipparchian fatherhood of the original work from which this list came drawn, and Otto

[2] According to Toomer (1980), Hipparchus was born in the first quarter of the 2nd century BC, and died after 127 BC.

Neugebauer, too, seemed later to give credit to this idea (1975). An anonymous commentator of Aratus (Maas 1898) reports also that Hipparchus said that there were 1080 stars contained in the constellations, but the origin of this statement was not known.

Another trace of the lost star catalogue of Hipparchus appeared in 1936 when Wilhelm Gundel found in two versions of the astrological treatise *Liber Hermetis Trismegisti*, one in ancient French, of 14th century, and one in Latin, of 1431, a list of 68 stars with longitudes corresponding to the time of Hipparchus. The terminology used, and the profile of the constellations, are similar to that of the *Commentaries*, although they do not contain ecliptic longitudes. Ptolemy cites some Hipparchian declinations (VII, 3), but it is not said that they were taken from the Hipparchus catalogue: for example, they might have been derived from one of his work on precession. Finally, in a fragment of Aratus composed of a macaronic Latin dating back to the 8th century, the so-called *Aratus latinus* (Maas 1898), values of polar distances and longitudes were found for the circumpolar constellations, approximately correct for the age of Hipparchus.

## 3 THE *COMMENTARIES*

Commentaries have been handed down, in whole or in part, in about fifteen codes, dating from the 14th to the 16th century, with the exception of one from the 11th century (Manitius 1894). In four of them there is a title, whose common part reports in Greek substantially: Ἱππάρχου τῶν Ἀράτου καὶ Εὐδόξου Φαινομένων Εξηγήσεων, that is *Commentaries by Hipparchus on the Phenomena of Aratus and Eudoxus*. The first printed edition of this work, in Greek, was published in Florence in 1567 by Piero Vettori, together with an edition of the Ptolemy's star catalogue that Vettori attributed erroneously to Hipparchus, and with a comment on the *Phenomena* by Achilles Tatius. The first Latin translation was published in Paris in 1630 by Denis Pétau in his work *Uranologion*, along with the commentary by Tatius, the *Introduction to the phenomena* by Geminus, the *Phases of the fixed stars* by Ptolemy, and other works. A second edition of the *Uranologion* was published by Jacques Paul Migne in 1857 in Paris, as an appendix to the XIX volume of his *Patrologia Graeca*. Moreover between 1888 and 1898 Ernst Maass published all the commentaries on Aratus that came to us, so also the first book and part of the second book of Hipparchus. The first translation into a modern language of *Commentaries* was the one in German published by Karl Manitius in 1894.

The work is divided into three books. The first book and the first part of the second are dedicated to the commentary and to a severe criticism not only of Aratus' *Phenomena*, but also of those of Eudoxus and of the commentary on Aratus, now lost, made by the mathematician Attalus from Rhodes. The second part of the second book and the third book are about Hipparchus' exposition of the risings and settings of the main constellations for a latitude of 36° N.

## 4 THE *COMMENTARIES'* DATING

According to Toomer (1980) the *Commentaries* are not an early work of Hipparchus, but reveal a mature astronomer. Therefore, he believes that they are probably following the lost star catalogue and that the star coordinates of the *Commentaries* have been taken from the catalogue itself. On the contrary, Neugebauer (1969) believes that the catalogue by Hipparchus was not only following the *Commentaries*, but also following the discovery of the precession of equinoxes and, for this reason, the catalogue contained the stellar coordinates expressed in the form of longitude and ecliptic latitude (unlike in the *Commentaries*): since for precession the latitudes remain unchanged and only longitudes vary, the same catalogue can be conveniently used even in subsequent centuries simply by adding the constant of precession to the longitude of each star.

The astronomical coordinates given by Hipparchus are sufficiently precise to date the *Commentaries*. For this purpose, I used the software *Starry Night Pro Plus* which is very reliable on both precession and stellar proper motions. I have used the coordinates of 96 stars expressed in terms of declination, polar distance or right ascension, looking for the year in which they best matched to the real one. These data give an epoch centred on 137 ± 8 BC (the uncertainty is the standard deviation of the mean). By excluding the most extreme values and considering those within Hipparchus's lifetime (56 data) and activity (38 data) respectively, we find 152 ± 3 and 147 ± 2 BC respectively. These findings are very similar to those by Heinrich Vogt, who in 1925, using the coordinates of 77 stars close to the ecliptic or the celestial equator, found a date of 151 ± 4 BC, but excluding 10 stars with an error of more than 65 years from the average result. Nadal and Brunet in 1989, using the coordinates of 78 stars obtained by cross-checking their rises, culminations and sunsets, found a value of 141 ± 25 BC. Based on these independent analyses, it seems possible to assign the *Commentaries* a date between 140 and 150 BC, thus confirming it is not an early work, but one of Hipparchus' maturity.

If I try to date, with the same method, the 18 stellar declinations of Hipparchus reported by Ptolemy in the *Almagest*, I find a value of 122 ± 20 BC; excluding the five furthest from the mean I found a value of 132 ± 5 BC, very similar to that indicated by Vogt, 131 ± 6 BC, who used 16 declinations, excluding the four of the *Almagest* with an error of over 52 years but taking two from Ptolemy's *Geography* and Strabo. In all cases, there seems to be some time lag between the writing of the *Commentaries* and the compilation of the coordinates quoted by Ptolemy, and this seems to suggest that Hipparchus compiled his star catalogue significantly later the writing of the *Commentaries*. In fact Ptolemy must have taken these coordinates from a scientific work by Hipparchus, probably his own star catalogue, and certainly not from a text of literary criticism, albeit on a scientific basis.

## 5 DETAILED ANALYSIS OF SOME PASSAGES OF THE *COMMENTARIES*

Hipparchus tells at the beginning of the work what was the purpose that led him to write a comment on the Aratus *Phenomena*:

> ... one could consider very useful and scientifically opportune to understand the affirmations he sustained on the celestial bodies, and which of them were written in agreement with the celestial phenomena and which were not ... Observing therefore that Aratus turns out to be in contradiction with the phenomena and with the reality on many and essential points, ... I thought ... to expose what seems to me wrong. I proposed to do this, not in order to gain prestige by refuting others (this would be a paltry and petty thought; on the contrary, I believe that we must be grateful to all those who personally undertake labours for the common interest); but so that neither you nor others who are eager to know may be led astray in observing the phenomena of the cosmos. Which indeed has happened to many; for the grace of the poems lends a certain credibility to what is asserted in them, and almost all those who comment on this poet agree with his assertions.

A little further on, Hipparchus seems aware that for a poet it does not claim the absolute adherence to the celestial phenomenon, and implies that his criticism is mainly directed towards Eudoxus, the astronomer who inspired Aratus:

> Eudoxus, with greater competence, wrote a treatise on the same subject as Aratus ... Therefore, it is perhaps not right to attack Aratus, even if in some cases he commits some errors. In fact, he wrote the *Phenomena* following the treatise of Eudoxus, not having personally observed the sky or proposing personal astronomical considerations on celestial phenomena ... That ... Aratus followed the writing of Eudoxus on the phenomena can be understood by comparing in several passages ... the expressions in prose of Eudoxus with the verses of the poet ... Two books on the phenomena are attributed to Eudoxus, which are in agreement on almost everything, except a few points. One is entitled *Mirror*, the other *Phenomena*. On the *Phenomena* Aratus composed his poem.[3]

However, Hipparchus from here on seems to forget several times these praiseworthy intentions. In the next pages, we will show some examples of the criticisms that Hipparchus addresses to Aratus and Eudoxus. In order to evaluate them I used *Starry Night* software utilized for dating.

Subsection 5.1 in the text includes all the parts of Book I of the *Commentaries* that I have commented on, and subsection 5.2 includes all the parts of Book II of the *Commentaries* that I have commented on. The subsubsections of the text correspond (with the exception of the last) to chapters of the *Commentaries* as Manitius (1894) divided them in the only critical edition published so far, and Arabic numerals in bold followed by a dot within the subsubsections correspond to the paragraph numbers with which Manitius divided the work within the chapters. The indented paragraphs are those of the citations. Hipparchus' quotes from Aratus are recognizable because the number corresponding to the first verse of the quote is always reported. The quotations of Eudoxus by Hipparchus are either implicit or, when they are explicit, they are never separated but always enclosed in quotation marks. In square brackets, I specify the stars indicated by Hipparchus and added words to make the speech clearer and fluent. Between braces there are some additions by Hipparchus to the quotations of Aratus. The translation is based on the Greek text of the critical edition by Manitius and is part of the second modern language translation of the *Commentaries* (Vanin and Cusinato 2017[2]).

### 5.1 (First Book of the *Commentaries*)

#### 5.1.1 (Chap. IV)

> **2.** Then, they are all in error also in the placement of the Dragon, supposing that it makes a curve around the head of the Little Bear. In fact, the brightest and western stars among those which are in the quadrilateral of the latter, and of which

[3] Besides by Hipparchus, this debt of Aratus towards Eudoxus is attested by two of five *Vitae*, *I* and *III*, which contain the few scarce biographical news about the poet. They state that Aratus was induced by Antigonus II Gonatus, king of Macedonia, to put in verse Eudoxus, giving him the treatise of the astronomer. But *Vita I* affirms that it was the work entitled *Mirror*, not the *Phenomena* as claimed by Hipparchus, while *Vita III* argues among other things that Aratus didn't just copy Eudoxus, but also added his own (Martin 1974). According to *Vita I* Aratus flourished at the time of the 125th Olympiad, or between 280 and 276 BC, which is therefore reasonable to assume as the time of the composition of *Phenomena*. According to Martin, *Vitae*, at least in the parts partially or totally in agreement, have a common origin in a grammarian of the first century BC, Teone, lived in Alexandria, but we do not know in general when they had been composed. As for Eudoxus, according to the *Chronicle* of Apollodorus of Athens (quoted by Diogenes Laërtius, *Lives and opinions of eminent philosophers*), he flourished in the 103rd Olympics, or between 368 and 364 BC, which, in absence of further precise indications, is admissible as the time of writing of his works.

the northernmost [β UMi] is, according to them, on the head, the southernmost [γ UMi] on the forelegs, lie parallel very close to the tail of the Dragon. Therefore, the following is not correct:

52 Cynosura has its head in the coil; on the same
head, it turns.

So writes also Eudoxus.

Cynosura is another name by which the Greeks identified the Little Bear. In this case, we have a first example of how Hipparchus often attributes to Aratus and Eudoxus statements that they did not make: we do not know what Eudoxus wrote on this subject, but certainly Aratus never said that the star β Umi is on the head of the Little Bear. For Ptolemy, β and γ Umi will be simply two stars in the rectangle of the Little Bear, so also for Ulugh Beg, while al-Sufi will place them in the back, the Vienna Manuscript and Dürer in the shoulders, just to mention ancient catalogues and iconography or derived from them. In any case, the two stars mentioned by Hipparchus, whatever part of the body of the Little Bear they represent, are certainly much closer to the spiral than to the end of the Dragon's tail, represented by the star λ Dra (10° vs. 17°).

**8.** … But the circle always visible,[4] in Athens and surroundings, … is about 37° from the pole.

Instead, the latitude of Athens is almost 38°, 37°58’ for precision.

**5.1.2 (Chap. V)**

**1.** In the following statements about the Bear, I think they are completely wrong; Eudoxus when he says: “under the head of the Great Bear the Twins lie, at the centre the Crab, under the hind legs the Lion”; and Aratus:

147 Under the head the Twins, under the central part the Crab,
under the hind legs the Lion shines with a beautiful glow.

Attalus and all the others agree with them. **2.** But that they are wrong it is clear from this: in fact, the head of the Great Bear according to the mentioned authors is the northernmost of the two western stars [α UMa] that are in the quadrilateral, while the southernmost of the same stars [β UMa] is on the fore legs. **3.** In fact, that the star in question is on the head comes from the fact that they say the star at the end of the Dragon’s tail [λ Dra] faces the head of the Bear. **4.** There are no other stars under the one at the end of the Dragon’s tail, except for the northernmost of the western stars of the quadrilateral. In fact, the one which is at the end of the Dragon’s tail occupies the 3rd degree on the circle parallel to the Lion, while the aforementioned star in the quadrilateral lies at just under the 3rd degree of the

[4] The stars that are located to the north of the ever visible circle are circumpolar. For a given location, the radius of this circle is equal to latitude.

Lion. **5.** That on the fore legs lies the southernmost of the western stars of the quadrilateral, Eudoxus explains by saying: “There is a bright star in front of the fore legs of the Bear.” And Aratus:

143 so they move beautiful and bright in front of its legs
one in front of that [legs] under his shoulders, one in front of those who come down from the loins.

**6.** Only one bright star is to the west of the southernmost of the western stars of the quadrilateral, the one currently marked in the fore legs. Generally, all the ancients depict the Bear with seven stars only.

Here, Hipparchus pretends not to understand that Aratus and Eudoxus adopt the “big” version of the Great Bear, not limited to the seven stars of the Big Dipper, but extending to all those, 24, listed for example by Eratosthenes;[5] Ptolemy will list 27 (VII, 5, II). From this point of view it is true their affirmation that under the Great Bear on the spring evenings of today as of 2000 and even of 2500 years ago, come to be the constellations mentioned, since in this version the Bear extends more towards the west, towards the Twins, and the head is not represented by the star α, but by several others, the latter of which, to the west, is the ο, which outlines the tip of the muzzle. In this “extended” configuration of the Great Bear, the legs correspond to the stars ι, κ, λ, μ, ν and ξ UMa. Therefore, the stars standing in front of the legs, as Aratus says, are evidently α, β and γ UMa, and in front of them does not mean in this case to the west, as it seems to mean Hipparchus, but simply “facing”. Furthermore, Hipparchus’ assertion that all the ancients consider the Great Bear formed by seven stars causes many perplexities, either because he does not cite any author, or because he himself, when presenting his phenomena, will use without problems the extended representation: in fact, in II.5.2 he says that when the Crown ends to rise, with star ε, pass in meridian at the same time “the star clearly visible to the west of the head of the Hydra, in the southern legs of the Crab, and the northernmost of the two stars in the fore legs of the Bear”. And since, when ε CrB arises, the Crab star that passes in meridian is the β Cnc, that of the Bear must be ι UMa, and certainly not β (It should also be mentioned that in the aforementioned manuscript Cod. Paris. 2506 attributed to Hipparchus there are also 24 stars in the Great Bear). Then, it is not possible to understand why Hipparchus affirms that Eudoxus puts on the fore legs the southernmost of the western stars of the quadrilateral, when Eudoxus instead says “*in front* of the fore legs there is a bright star”, and it is not understandable to which star Hipparchus refers when he writes that “one only bright star is to the west of the southernmost of the western of the quadrilateral, the one that is currently marked in the fore legs”, because he himself, shortly before, says that the southernmost of the western of the quadrilateral is placed on the fore legs, and among other things to the west of this star, β UMa, there are no other bright stars.

**14.** Aratus on the Charioteer still writes:

177 But the Bull is always faster than the Charioteer
to reach the setting, even though it has gone up together.

[5] *Catasterismi*. Although *Catasterismi* that has come to us almost certainly are not in Eratosthenes own hand, and therefore is not preceding Hipparchus, is based on a work of Eratosthenes which was to be more comprehensive and exhaustive: see on the whole question Vanin and Cusinato (2017[2]).

But it seems to me that he disagrees with the phenomena even in these verses: only the feet of the Charioteer rise with the Bull, the rest of its body rises with the Fishes and the Ram.

Actually, the first part of the Charioteer rose along with the last part of the Ram (constellation and sign), the latter part along with the first part of the Bull (constellation and sign).

**15.** And he himself says in the following verses:

718 But the Kids and the sole of the left foot with the Goat itself
are grouped with the Bull.

But, according to him, all the parts that rise with the Bull rise with the Ram.

However, this appears only an interpretation of Hipparchus, which does not seem to be supported by Aratus' text.

**16.** Long before the left foot [of the Charioteer, ι Aur] rise the right shoulder [β Aur] and the right hand [θ]. It is clear, therefore, that, also according to Aratus himself, only the parts near the feet rise with the Bull, the rest of the body and even the left foot rise with the Ram.

This is also a subjective interpretation of Hipparchus.

**17.** And even this is not in his favour, because he does not speak of the Bull as a whole, but of the fact that the Bull sets before the parts of the Charioteer that have risen with it. This is also false: in fact, his feet not only do not follow the Bull, but rather set before. Specifically, the star on the right foot of the Charioteer [β Tau] sets with the 27$^{th}$ degree of the Bull. **18.** Therefore if he had written in accordance with the phenomena, as Attalus affirms, it would have been much better to say, and more remarkable, that the right foot of the Charioteer, risen later, sets before [of the Bull] and not that, risen together, sets afterwards.

However, Hipparchus' statement is false: even β Tau (which for the ancients was the right foot of the Charioteer, but also represented the tip of the northern horn of the Bull) set seven minutes after the last star of the Bull, ζ Tau (representing the southern horn). Even considering the sign and not the constellation, his statement is not true: in fact Hipparchus corrects himself just after, when he says that this star sets with the end of the Bull sign, the 27$^{th}$ degree (actually 25$^{th}$), not before.

**20.** In the followings passages Aratus is wrong, saying of Cassiopeia:

188 in front of him then turns unfortunate.

Because Cassiopeia is to the east of Cepheus.

And on the Lyre equally states Aratus:

270 put {this one, Hermes, says}[6] in front of an unknown figure.

Instead, it lies to the east of the Kneeler.

Hipparchus seems to interpret these "in front" as if they meant "staying west, western", or "staying in front" in the sense of the movement of the celestial sphere, which takes place from east to west. Instead, Aratus in these places simply refers to the reciprocal positions of the various figures.

**21.** In an incorrect way Aratus also says of Cassiopeia:

188 certainly not great
appearing on a full Moon night, Cassiopeia;
in fact not many stars in succession make her shine.

Indeed most of its stars are brighter than those in Ophiuchus' shoulders, which, also, he affirms, are well visible on a full Moon night, with these words:

77 such the bright shoulders placed under his head
you see; even those on a full Moon night
would appear visible.

**22.** They are almost brighter even than the stars of Andromeda, except the one in the head and the eastern of those on the belt [β And], about which he says:

198 I don't think you much
at night have to look for, to see it immediately well,
so [visible] is the head, such on both sides
the shoulders and the ends of the feet and the entire waistband.

Here, Hipparchus misunderstands Aratus in the sense that he thinks that in verse 188 πολλὴ means "shining", whereas Aratus means to refer to the size of the constellation. This error has been made by several translators, but Kidd (1997) has convincingly shown that Aratus uses πολλὴ and πολλός in the description of stars and constellations to indicate size, not brightness (vv. 87, 165, 255, 316, 611, 699). Especially emblematic is the case of Capella (verse 165) where he clearly distinguishes the apparent dimension (πολλὴ) from luminosity (ἀγλαή). Certainly Aratus in verses 188-190 played with words a bit (as a poet must do...), but Hipparchus had to understand, from a minimal comparison with the other places, what sense Aratus gave to the word πολλὴ.

### 5.1.3 (Chap. VI)

**1.** It is also found in Eudoxus an erroneous representation of the head of the Great Bear, … in the treatise entitled *Phenomena*, as follows: **2.** "Under Perseus and Cassiopeia not far away there is the head of the Great Bear; the stars in the middle are weak"; in the *Mirror* [he says] instead: "behind Perseus and not far from the hips of Cassiopeia there is the head of the Great Bear; the stars in the middle are weak." **3.** But the head of the Great Bear is not in the region of Cassiopeia and Perseus, and it's not little far from it; in fact Cassiopeia lies above the twelfth of the Fishes, Perseus above the Ram, the head of the Great Bear, according to Eudoxus, about in the 2nd degree of the Lion.

[6] The part in brackets is added by Hipparchus; it should be understood: "Aratus says that Hermes put the Lyre...".

It does not make sense, for constellations near the pole, to refer to their angular distances in the sky by quoting only, as Hipparchus does, their right ascensions. In fact, the distance between the head of the Great Bear, represented by the stars on its forehead, ρ and σ UMa, and the nearest star of Cassiopeia that is sufficiently bright, ι Cas, is only 34°, not 137° as malignantly Hipparchus suggests. Certainly Perseus, Cassiopeia and the Great Bear are not very close, but the "not very far" of Eudoxus seems quite appropriate. Among other things, it is important to remember that in the ancient times there was no constellations in the intermediate space: Camelopardalis and Lynx were introduced only in the 17th century.

**8.** Aratus also supports unfoundedly the following:

239 They still farther ahead, and moreover on the thresholds of Notus,
the Fishes.

In fact, they are not both more southern than the Ram, but just one of them. The [stars] on the muzzle of the northernmost Fish [82 and σ], which are somewhat more southern than the eastern star [β And] of those in the Andromeda belt, are distant from the northern pole 70°; the western star of those in the tail [η And] is distant from the northern pole 78°. **9.** Of the stars in the Ram the northernmost and on the muzzle [α Ari] is distant just under 78°, just like the northernmost [π] of the stars in the tail; the stars in the Ram's body are all more southern than these. It is clear therefore that one of the Fishes is more northern than the Ram.

Most of the stars of the Fishes are south of the Ram. However, besides that, Hipparchus' criticism is doubly puzzling. First, commenting on the same verses shortly before (I, 5, 20) he states that "ahead" in this case means westward, and then Aratus says, as a matter of fact, that the Fishes are both west and south of the Ram, as it is in the reality, and not only to the south, as Hipparchus pretends to understand. Moreover, unlike Aratus, Hipparchus thinks about the zodiac in terms of zodiacal signs, not constellations, and from this point of view the Fishes are a 30° ecliptic portion that lies definitively south of the Ram.

**14.** Not exactly Aratus also says that the Pleiades include only six stars:

261 Seven {in fact he says} are those indicated by a name,

258 although only six are visible.

He does not say though that those who look intensely on a clear night and without Moon see seven stars in the cluster.

The visibility of the seventh star has always been a never-ending story from the earliest times, so that it has fed the famous legend of the lost Pleiad. A person with normal vision sees six stars in the Pleiades, which are Alcyone, Atlas, Electra, Maia, Merope, Taygete. Those with acute eyesight can usually see not only a seventh, but also an eighth and a ninth star. It seems that the pre-telescopic primacy belonged to Michael Maestlin, who saw 10, 11 or even 14 (*Historia coelestis* 1672; Winnecke 1878; Kepler 1610).

### 5.1.4 (Chap. VII)

> **19.** ... both wrong, Aratus and Attalus who agrees with him, on the fact that Cepheus sets with the waist:
>
> 650 he grazes the ground, the parts near the head all
> plunging into the Ocean; to the others, this is not allowed,
> to the feet, to the knees, to the side: the Bears prevent it.
>
> **20.** At the latitude of Greece, Cepheus does not plunge to the waist, nor to the shoulders. Only the stars in his head come down just under the horizon; the shoulders move in the always visible arch, without rising or setting; in fact the bright star in the right shoulder [α] is 35½° far from the pole, the bright star in the left shoulder [ι] is at 34¼°. **21.** Where the longest day lasts 14.5 hours, there the circle always visible is 36° far from the pole, in Athens 37°. It is clear, then, that the bright stars on Cepheus' shoulders ... move ever further north of the circle always visible.

The Hipparchus' citation is incomplete, and must be integrated with the second part of verse 649 ("...then Cepheus with the waist"): therefore, Aratus does not say that Cepheus sets with the waist but that with the waist it grazes the ground, two very different things, and that only the parts close to the head set. In verse 310, he says that it sets to the side and at 633 "with head, hand and shoulder". At the time of Eudoxus the head sets completely, together with his right hand and actually also the right shoulder was invisible because of extinction, reaching a height of only one degree (with refraction). At the time of Hipparchus, the variation was minimal and the stars were only 40' higher. Observing from Rhodes the variation would have become even smaller, imperceptible to the naked eye, just over a tenth of a degree. Hipparchus' criticism is partially meaningful only if a celestial globe is used, but it is unmotivated when referred to the real sky.

### 5.1.5 (Chap. VIII)

> **1.** Aratus is also mistaken about the constellation Argo. He says in fact that the part of this going from the bow to the mast is devoid of stars. His words:
>
> 349 Partly all indistinct and starless as far as the actual
> mast from the prow turns, partly all bright.
>
> In fact, the bright stars placed in the cut of the ship, of which the northernmost [κ Vel] is in the bridge, the southernmost [ι Car] in the keel, are located far eastwards.

Aratus here undoubtedly relies on the tradition that Argo is seen in the sky only in the part that goes from the stern to the mast (Eratosthenes, 35; Vitruvius, IX, 7; Hyginus, II, 37; Ptolemy, VIII, 1, XL). So, probably, for Aratus, the cutting of the ship coincided with the mast's position. However, the two stars could be seen at the time of Eudoxus, but ι Car was hardly visible, only for a few tens of minutes, at the culmination, with an elevation of at most 2°44′ in Knidos, where Eudoxus lived.

**14.** Afterwards Aratus says about the Censer:[7]

402 But beneath the flaming sting of the great monster,
the Scorpion, near Notus, hovers in the air the Altar.
But this, indeed, for a short time high
you will observe; in fact it rises on the other side of Arcturus;
and while are very high in the sky the paths of
Arcturus, it sooner plunges into the western sea.

In these verses it seems to me that Aratus is wrong, thinking that as Arcturus is far from the pole always visible, as much is the Censer from the southern pole. **15.** Similarly Attalus is wrong, because he agrees with him. Explaining in fact the meaning of the above verse, he says: "speaking of the Censer he states that it bears the same relation to the invisible pole as the star called Arcturus does to the visible pole. So he says that the motion of the Censer above the earth is short, while that of Arcturus is long". **16.** But they are wrong in thinking that Arcturus has the same distance from the northern pole as that of the Censer from the southern pole. First of all, in fact, the stars in the Censer do not lie on the same parallel, but among them some are much more southern, some more northern; if then, apart from this, we make the comparison with the centre of the constellation, nor so the statement would be valid; in fact, Arcturus is 59° from the northern pole, the bright star at the centre of the Censer [α] is 46° from the southern pole.

Again, Hipparchus, this time followed also by Attalus, attributes to the poet intentions that he does not have. Aratus only states, according to reality, that the Censer rose from the opposite side where the star Arcturus sets (at the time of Eudoxus the azimuth difference was about 163°) and while the latter followed very high trajectories and remained visible for a very long time, the Censer was seen for a few hours, setting rather quickly. Among other things, the star α is not in the centre of the Censer, as Hipparchus says, but on its northern edge.

**23.** There is also an error in the verses immediately following the preceding ones:

439 On the other hand he looks like one who always tends the right
against the turned Altar.

In fact, between the right arm [η Cen] and the Censer lies the whole Beast and most of the body of the Scorpion. The right hand is placed around the 8th degree of the Claws, and the Censer [lies] under the extreme parts of the Scorpion, as Aratus himself says …[8]

[7] Hipparchus uses a different term than Aratus and Eratosthenes, θυμιατήριον, literally just "censer, thurible", followed by Geminus and Ptolemy. Obviously where he quotes the verses of Aratus uses the term employed by him, θυτήριον, "altar".

[8] At this point in the text Hipparchus still quotes verses 402-03.

Hipparchus appears to be still too pedantic: the Centaur's right arm is anyway stretched towards the Censer, although on the right lies the Beast, of which Aratus speaks immediately afterwards, and although the Censer is a bit detached from the Centaur. Moreover, the Scorpion is not at all interposed between the two constellations, but it is entirely north of the Censer, as on the other hand Hipparchus recognizes immediately afterwards.

### 5.1.6. (Chap. IX)

> **14.** After talking about the milky circle he adds that, of the four circles, two are equal, two much smaller:
>
> 477 and indeed no other circle similar in colour to this
> turns, but equally large in size on four there are
> two, the others, much smaller than these, rotate.
>
> **15.** It does not seem to me that this is even true, that the tropics are much smaller than the equinoctial and zodiacal circles; in fact they are smaller than less than $\frac{1}{11}$.

Another pedantic and basically useless observation: as a matter of fact, $\frac{1}{11}$ may seem small, but it depends on what you compare. For example, if you instead of eating 100 g of spaghetti you eat 91, it is difficult to notice the difference; but if a man is 1.64 m, he is just $\frac{1}{11}$ shorter than one who is 1.80 m: how would we define him, a little shorter, or much shorter?

### 5.1.7 (Chap. X)

> **19.** And then he says:
>
> 518 In it the belt of the shining Orion
> the curve of the inflamed Hydra, in it also the faint
> Bowl, in it the Raven.
>
> Therefore, the Orion's belt lies on the equinoctial circle, but the Dragon's[9] coil and the Bowl and the Raven are much further south of it, with the exception of the area around the Raven's tail, which approaches it.
>
> 520 in it, {says},[10] the not many stars
> **20.** of the Claws, in it, are the knees of the Ophiuchus.
>
> But of the Claws only the bright star in the northern Claw is close to the equinoctial circle, the others are much further south of this.

The Orion's belt has never been on the celestial equator during the current precessional cycle. At the time of Hipparchus the belt was between 4°30' and 5°40' south of the equator (at the time of

[9] Hydra, of course. Hipparchus' oversight.
[10] Addition by Hipparchus.

Eudoxus still 40′ further south), so it is unclear why Hipparchus did not detect this error. On the other hand, it is really hard to think that the astronomer of Rhodes recognized in other stars the belt, which was catalogued by Ptolemy exactly as we see it. Moreover, the curve of the Hydra, distinguished by the stars ι, $\tau^1$ and $\tau^2$, was north of the equator at the time of Hipparchus, and not south. Finally, even though the equator did not pass just in the middle of the Claws, however crossed a good portion of the northern part of it, and therefore Aratus' statement is in good agreement with the phenomena.

### 5.1.8 (Chap. XI)

> **7.** ... it is not correctly said that the star called Canopus moves in the invisible circle:[11] this is in fact the southernmost bright star of the [Argo] rudder. It is distant from the south pole about 38½°. **8.** The always invisible circle in Athens is about 37° from the pole, about 36° in Rhodes. It is clear then that this star is further north of the invisible circle in Greece and can be seen moving over Earth. And certainly you can see it in the places around Rhodes.

The distance of the star Canopus from the south pole at the time of Hipparchus was 37°18′ and not 38½°, too little, even taking into account the refraction, to be seen culminating in Athens, that is at almost 38° latitude. The observation was also impossible at 37° latitude. In fact, it should be remembered that atmospheric extinction greatly weakens the light of the stars low on the horizon. Even counting the refraction, Canopus' elevation at the culmination, at 37° parallel, was only 49′, and at that elevation the average extinction is 7.9 magnitudes. Therefore, the magnitude of the star, -0.6, would become 7.3, invisible to the naked eye. Also taking into account the very clear sky of winter nights, in which the star reached the culmination, with a lower coefficient of atmospheric absorption, the extinction could not be less than 6.9 magnitudes, and so Canopus could not exceed the threshold of visibility at naked eye (Green 1992). On the other hand, we have the witness of Geminus who, about a century later, says that even in Rhodes (latitude of the capital city of the island 36°26′) Canopus could only be seen in very clear nights and/or high places (*Introduction to the phenomena*: III, 15. At the time of the writing of the Geminus' work, about 60 BC, the distance of Canopus from the pole was basically the same as for Hipparchus, 37°21′). Among other things, Eudoxus does not say, as Hipparchus attributes to him, that the star "moves in the invisible circle" but only that it is "very close to it", a remark that is probably the result of an experimental consideration. In fact, Posidonius writes that Eudoxus, in his observatory at Knidos, in a higher position, but not much, than the surrounding houses, could see Canopus at the culmination (Strabo: II, 5), an observation in line with the theoretical expectation: being Knidos, at 36°41′ lat. N, the star, which at his time was 37°10′ from the celestial southern pole, taking into account the refraction reached at the culmination an elevation of 58′. Thus, it was just visible in the clearest nights, appearing of magnitude 5.9, due to extinction. Posidonius himself confirms that under normal circumstances Canopus could not be seen from Greece (Cleomedes, I,10) and the northernmost place where the observation is possible is Rhodes (which he assumes on the other hand to be on the same parallel as Knidos, Strabo, II, 5).

---

[11] The stars that are located to the south of the always invisible circle are not ever visible from a given latitude. Also the radius of this circle is equal to the latitude of the site.

> **19.** He is also mistaken[12] to say that his body is cut in half by length by the circle; in fact, the star on the head [β Boo] is about 16½° of the Claws, the star in the right foot [ζ Boo] lies about 24¾° of the Claws, the bright star in the belt [ε Boo] is located around 14⅓° of the Claws. **20.** ... The head of the Sea Monster is to the east of the circle not by much; ...

Here, speaking of the stars' positions within various constellations on the equinoctial colure, Hipparchus makes several errors: the β is at 25°, the ζ at 14°, the ε at 17½° of the Claws. Certainly, these three values may have been swapped by an inattentive amanuensis, but in the last sentence there is another remarkable dislocation, since the Sea Monster's head at the time of Hipparchus was east of the circle by almost 14° (the colure bisected its head in 1250 BC).

### 5.2 (Second Book of the *Commentaries*)

#### 5.2.1 (Chap. I)

> **2.** First, so Aratus says, wishing to teach how one can know the hour at night, through the rising and setting of the constellations:
>
> 559 It would be profitable for those who are waiting for the day
> to observe when each part rises; …

In the section of his poem that runs from verse 559 to verse 732, Aratus deals with the simultaneous rising and setting of the constellations and, speaking of the zodiac, he uses several times the word μοῖρα, "part", which is the technical term used by the Greeks for "twelfth part of the zodiac", and therefore it may seem that he refers to the signs, not to the zodiacal constellations. In reality, in the whole poem, Aratus refers to the constellations, as it is evident from repeated references to parts of things, humans and zodiacal animals, for example, in verses 89, 97, 137, 147-48, 167-78, 232, 238-47, 282-86, 386-92, 402-03, 438, 446, 481, 491-95, 500-02, etc., and even in this section, he continues to refer to the constellations, as itis quite clear in 563, 572-73, 715. However, it could not be otherwise, because Aratus describes the real sky and therefore he cannot talk about anything else than what is seen directly in the sky. The zodiacal signs are a partition of the ecliptic in 12 equal extension portions, 30° each, and have no counterparts in the real sky, and they can only be identified on a map or a celestial globe. Nevertheless, in the sequel, Hipparchus blames Aratus because he considers the zodiacal constellations, which have a different extension to each other, and not the signs, which have constant extension, and this approach would not provide a precise hourly indication. Apart from the fact that this latter statement is not true at all, since it would be sufficient to rely, within each zodiacal constellation, on certain stars or groups of stars, as it was done with the Babylonian culminating stars or the Egyptian decanes (or by Hipparchus himself with his hourly stars, though not all zodiacal), it is clear from the quoted verses that Aratus is not interested in observing the rising and the setting of the zodiacal constellations to infer the hour at night. He is interested instead in their rising to know when the Sun will rise. In fact, in this section, he limits himself to associate the zodiacal risings with the risings and settings of the other constellations.

[12] Always Eudoxus.

Despite this, Hipparchus, over pages and pages, makes a severe but perfectly useless criticism, opposing to the risings of the real sky of Aratus his risings based on equidistant zodiacal signs traced on a celestial globe. Furthermore, Hipparchus criticizes Eudoxus, showing that, although he considers, for the simultaneous risings, the signs and not the constellations, he makes a lot of mistakes. However, he does not offer any proof of the adoption of the signs by Eudoxus, while in some of his quotes from Eudoxus himself, in which parts of constellations are cited, we have the opposite proof: "the "body of the Lion" (I, 2, 18), "the feet of the Water-pourer" and "the sting of the Scorpion" (I, 2, 20), "the back of the Ram" (II, 1, 22). Finally, it is not necessary to overemphasize this aspect. Therefore, I will not recall most of this criticism in the continuation, limiting myself to point out only inaccuracies in the risings and settings of the same signs as calculated by Hipparchus, or inconsistencies of other nature.

**5.2.2 (Chap. II)**

> **11.** About Arctophylax it seems to me that they are completely wrong; in fact they say that this one sets in opposition with four zodiacal signs, the Ram, the Bull, the Twins, the Crab. **12.** In fact, Aratus writes that when the Bull is rising:
>
> 721 Sets Arctophylax already with the first part
> of the four that drag him down, excluding his hand.
>
> So he begins to set in opposition with the rising Ram.

Here, Hipparchus seems to contradict himself: first, he writes that Aratus assumes that the first part of Arctophilax sets when the Bull is on the eastern horizon, and then asserts that this happens with the appearance of the Ram. Incidentally, the left foot [η, τ and υ Boo] sets with the Pleiades well high, while Arcturus sets with the Hyades already risen. However, it should be noted that here and in other places Hipparchus plays on the ambiguity of the various phenomena described by Aratus, in the sense that when he writes that a constellation rises (and the same applies to sets) it is not entirely clear whether he intends that it is about to rise (and therefore no star is still visible), that it begins to rise (and so the first stars are visible), whether it is risen half, three quarter or whole. And Hipparchus, instead of honestly acknowledging the correct approximation of every poetic writing, here and in many other parts arbitrarily establishes certainties absolutely inopportune.

> **35.** The whole Dog rises together with the Crab, except the star in the tail [η], and not, as Aratus says, only its forelegs [β].

But the β appeared when all the stars of the Crab were already abundantly risen (the 15$^{th}$ of the Crab as a sign). With the Dog all risen, half-constellation of the Lion was above the horizon (5½° of the Lion as a sign).

> **38.** The River starting from Orion does not begin to set in opposition to the Lion, as Eudoxus says, but to the Maiden; in fact its western and southernmost star [θ], which is also the brightest, sets at rising of 7$^{th}$ degree of the Maiden.

Here, Hipparchus is wrong by 11° because the θ sets at the rising of the 26$^{th}$ degree of the Lion. On the other hand, to his partial excuse, this star culminated at a maximum height of 4° even in Rhodes, and therefore the measurements of its position as it set were heavily affected by refraction and extinction.

From par. 39 of chap. II to par. 35 of chap. III there are many Hipparchus' misinterpretations of what written by Aratus. Here are some examples from chap. II:

> **39.** He affirms that only Argo's stern rises together with the Lion, ...

While Aratus does not quite mention the Lion about Argo.

> **44.** The continuation of the speech is such: when the Maiden begins to rise, the Dog and Argo's stern have already risen; …

And even here it does not seem that Aratus says this.

> **46.** He says that together with the Claws ... they set ... and most of the River.

This is an inference of Hipparchus: about the River Aratus mentions the Scorpion.

> **47.** ... But regarding the Centaur they are wrong; in fact are not its tail and in general the back parts to rise first, but the left shoulder [ι]; it is in fact much more northern, and does not begin to rise together with the Maiden, as they argue, but with the Claws; …

But that is what Aratus also says, in 625-626.

> **50.** Of Cepheus only the head sets; his shoulders lie in the ever visible part, as we have already said. Not only they are mistaken about this, but also when they say that his head sets in opposition to the Maiden; ...

But Aratus says that Cepheus sets with the Claws and the Scorpion.

> **51.** Aratus asserts that when the Scorpion begins to rise set … Cassiopeia whole, except the part going from the feet to the knees.

This too is a supposition of Hipparchus, (badly) interpreting verses 654-656.

> **56.** Aratus then declares that when the Archer is about to rise, the head and the left arm of the Kneeler have already risen, so also Ophiuchus' body, and the tail of the Snake,...

Instead, Aratus (672-673) says that head and left arm of the Kneeler *rise*, not *have already risen*, together with the Archer, and in 665 he speaks of the Snake's coil, meaning probably the central one, not of the tail.

> **57.** ... he says that the Goat and the Kids, of which one on the left shoulder, the others in the left hand, moreover the head and the right hand, set in opposition to the Archer, while the feet set in opposition to the Scorpion.

Aratus, however, in verses 673 and 679-85 says that when the Archer rises, the Goat and the Kids are not yet completely set, while the feet go down with the Archer and the rest of the constellation with Capricorn.

> **59.** ... so that Cepheus' head not only rises together with the Scorpion, but also with the Archer. **60.** Whole Perseus sets in opposition to the Scorpion and not, as they say, the right foot and the right knee are brought down in opposition to the Archer.

Here, Hipparchus should have written the opposite: "not only with the Archer, but also with the Scorpion", because this is what Aratus says. Then again he seems to be confused, because shortly before he has reported correctly what Aratus has written, that is, that the right foot and knee *do not* plunge with the Archer (therefore, rather, with Capricorn).

### 5.2.3 (Chap. III)

> **4.** He says that when the Water-pourer begins to rise have already risen with Capricorn the head and the legs of the Horse, ...

But Aratus speaks of "central part of the Water-pourer" (see v. 693), as Hipparchus points out in par. 6. Similarly to other parts of his work, Hipparchus attributes non-existent statements to those he criticises, and once again he shows a certain carelessness in the criticism itself. The phenomena relating to Capricorn and the Horse are, in fact, congruent with the rising of the central part of the Water-pourer, not with the beginning of its rising, as Hipparchus attributes to Aratus.

> **5.** Undoubtedly, what they said in common [Aratus and Eudoxus] almost accords with the phenomena, except that the head of the Hydra begins to set in opposition with the last parts of the Archer, and not with Capricorn.

But Aratus does not say that the Hydra sets with the rising of Capricorn (vv. 693-98) but with that of the Water-pourer, and he speaks of head and coil of the neck, not only of the head. The last star of the coil, ι Hya, sets when the $12^{th}$ degree of Capricorn sign rose, had risen half-constellation of Capricorn and already the first stars of the Water-pourer (ε, μ and ν).

> **15.** On the Centaur both are wrong. It is not in fact completely set, in opposition, when the Fishes begin to rise, but its front parts are still above the horizon: in fact the head [1, 2, 3 and 4 Cen], and the right shoulder [θ], set in opposition to the Fishes.

We do not know what Eudoxus wrote, but Aratus, in verses 700-01, simply states that the Centaur sets when the Fishes rise, a general remark, as a poet, which is however correct and also perfectly congruent with what Hipparchus writes in the first part of his statement. Hipparchus, however, contradicts himself in the second part. In fact, if it is true that the right shoulder and the head of the Centaur have not yet set when the Fishes begin to rise, it cannot be said that these parts set in opposition to the Fishes: when two constellations are in opposition, i.e. when they are 180° apart in celestial longitude or 12h in right ascension or, more generally, when they are in opposite positions to each other, on two opposite sides of the horizon, one must refer to the same parts of the constellations, for example the beginning, the middle or the end of each one, not to different parts.

> **16.** Then they both disagree with the phenomenon also on the Southern Fish, assuming that it rises as a whole almost together with the Water-pourer. On the contrary the greater part of it rises with the Fishes ...

We do not know what Eudoxus writes, but Aratus says that the Fish (701) rises after the Fishes, which is correct from the astronomical point of view.

> **34.** They are mistaken, however, thinking that only the left side of the Charioteer rises together with the Ram; …

To tell the truth they have talked about the Bull, as just Hipparchus has said shortly before. Aratus even says that the Charioteer ends to rise with the Twins (vv. 716-17).

> **35.** Likewise, they were wrong about the descriptions of the Sea Monster, in two distinct ways. In fact it does not begin to rise, as they say, with the Ram, but with the Fishes, ...

But also here Aratus speaks of the Bull (vv. 719-20) and says it completes the rising with the Twins (vv. 726-27).

### 5.2.4. (mistakes in signs)

Finally, there are a number of mistakes that Hipparchus does, even taking into account the fact that he speaks of the signs. For example, in Book II, chap. II:

> **48.** Of Andromeda not only has set her head when the Claws are about to rise, but also both arms.

While Andromeda has set to the belt and beyond.

> **49.** ... and certainly the rest of the body of the Sea Monster does not set in opposition to the Maiden as a whole, but only to the ridge, as Aratus says.

But the last star of the Sea Monster set in coincidence with the rising of the very first stars of the Claws.

> **54.** ... But the Beast does not rise only with the Claws, as Aratus supposes, but also with the Scorpion: in fact it begins to rise when the 21$^{st}$ degree of the Claws is brought up, the rest undoubtedly rises together with the centre of the Scorpion, as also Eudoxus states.

In reality, the Beast rises with the first part of the Scorpion, both sign and constellation.

**61.** Also their observations about Argo are wrong: in fact it begins to set in opposition not to the Scorpion, but to the centre of the Claws. It had then to be said that Argo's stern has set when the Scorpion begins to rise, and not when the Archer rises.

Taking for the beginning of the setting of Argo that of Canopus (which for Hipparchus was obviously visible also from 37° latitude, having assigned to it a polar distance of 38.5°), the constellation of the Maiden had yet to finish rising on the eastern horizon, while there was the border between the Maiden and the Claws as a sign. However, for the latitude of 36° the constellation of the Maiden had completely risen, while the $9^{th}$ degree of the Claws was rising as a sign. This clarification is necessary because at the beginning of the second part of the *Commentaries*, where Hipparchus presents its *own phenomena*, he innocently confesses to expose them for a latitude where the day lasts 14.5 equinoctial hours, or 36°, not 37°. Since it is unlikely that he possesses two globes for the two different latitudes, he implicitly admits that the criticisms of the first part are subject to a not declared source of error, a behavior not scientifically correct.

## 6 CONCLUSIONS

Surely, the reading of the first part of Hipparchus' work leaves a lot of perplexity for the overly critical level towards the commented authors. Delambre (1817), the only one who in the past has tried to carry out an in-depth study of the work, translating or paraphrasing a large part of it, stated that he did not find anything bitter and unjust in these criticisms and tended to give reason to Hipparchus, thinking that Eudoxus did not have astronomical tools or that he limited himself to talking about things done by others. On the other hand, the many medieval scholiasts of Aratus generally defended the poet, excusing him in various ways, through myth, invoking erroneous interpretations, convoluted explanations (see for example Fedeli 2001), but without much foundation or proper astronomical investigation. It is precisely this type of investigation carried out here, that shows how the criticisms of Hipparchus are often too severe, pedantic and inopportune, his reliefs unnecessarily obstinate, and, worse than this, in many points he forces the authors' thinking by attributing things they have not said. Moreover, Hipparchus' criticisms are mostly wrong both when they apply to general phenomena and when they concern more precise aspects of positional astronomy. A simple quantitative analysis shows that only 41% (37 out of 91) of criticisms raised at Aratus, and only 47% of those against Eudoxus (26 out of 55), are astronomically motivated. In addition, the treatment is poorly linear and homogeneous, and sometimes we see real Pindaric flights between the topics, which make it difficult to follow his reasoning.

It is also possible that in some cases the criticisms are due to the fact that the constellations in different ages had different shapes. Already Gundel (1936) noted that there were several differences between the descriptions of the constellations contained in *Liber Hermetis Trismegisti* and those of *Almagest*. On the other hand, Ptolemy himself (VII, 4) states that his description, "more natural and proportionate", is different from that of his predecessors, as that in turn is different from even older ones. However, this does not seem to justify the severity of Hipparchus. Instead, perhaps the opposite is true. It should be noted that, apart from perhaps three or four cases, at the basis of Hipparchus evaluations there are no difference in the stellar positions between the ages of Eudoxus and Aratus and that of Hipparchus due to precession, since the statements of Aratus and Eudoxus are rather generic, are not aimed to have great accuracy, and the vast majority of the concerned phenomena have not changed significantly between the two epochs.

The second part of his work is certainly more enjoyable: whilst previous authors provided only lists of rising and settings of stars and constellations, Hipparchus provides the longitudes of the ecliptic points that rise, culminate and set, and which stars pass in meridian when various stars in the beginning and end of each constellation rise and set. He also provides, at the end, a list of bright stars lying on or near the 24 hourly circles to determine more precisely the time at night. It is reasonable to assume that the observation on rise and set as well as the observations of the phenomena of the first part have been done theoretically, using a celestial globe. Indeed, the numerous references to the degree of each sign that rises, sets or passes in meridian are not directly linked to the real sky, since the partition of the ecliptic in 12 equal parts is an artifice that can be traced on a globe but has no counterparts in the sky. Just as the references to the other circles are certainly better viewed and positioned using a celestial globe, accurately built and sufficiently large.

Instead, it is likely that the hourly stars of the second part and the numerous stellar coordinate values scattered in the first part are the result of direct observation. The coordinates used by Hipparchus are in the form of both polar distances, declinations and right ascensions. The latter are given in terms, however, unusual: for example, he says a star "occupies the third degree of the Lion along its parallel circle" (see above: I, 5, 4). This means that the equator, and each circle parallel to it, is divided into 12 "signs" of 30° each: so the right ascension of the above-mentioned star is 123°, or $8^h12^m$.

The values of the polar distances, the declinations and the hourly stars are quite accurate: an analysis of 67 "exact" data (without the "about" specification) shows that the average error is ⅓°, more or less that of the *Almagest* star catalogue. Most data on rising, setting, meridian passages and other phenomena is affected by an uncertainty of at most half degree or one degree. Hipparchus' largest errors, one or two degrees, are mainly for phenomena on the horizon, and are due to the ignorance of the refraction phenomenon.[13] When they are even larger, they are usually related to low declination stars, badly observable due to the low elevation they reach on the horizon. When there are noticeable errors on the meridian passages, they are related to phenomena on the horizon of high declination stars, whose path appears almost parallel to the horizon and of which is thus difficult, even with a globe, to evaluate the moments of rising and setting. There are, however, several macroscopic errors, of the order of 5° and more.

In conclusion, the second part of the *Commentaries* is not so free from mistakes and so good to erase the negative impression of the first part, a commentary that appears, to an astronomical inquiry, as a poorly-founded and bad-documented criticism, steeped of ill-concealed malevolence and acrimony. All in all, the Hipparchus emerging from a careful analysis of the *Commentaries* seems very different from the scientist which Pliny defined as "the confidant of the nature projects", "never praised enough", "admirable scholar" (II: 53-54, 95 and 247), that Plutarch denominated "high quality scholar" (*De facie quae in orbe lunae apparet*, 4) and that Ptolemy called several times "lover of truth" (III, 1 (twice) and IX, 2).

Perhaps, a more thorough examination by historians of his only remaining work could offer a more balanced judgment of both the figure of Hipparchus and the whole Hellenistic astronomy. Failing that, we risk judging Hipparchus' stature not for what he has written, but for what he has not written, that is, in essence, for the references and the judgment of Ptolemy, which may not have been a reliable judge. It might be noted that this would not be the only case in the history of science: for instance, Newton attributed to Galileo the discovery of the first two principles of dynamics, while the real merits of the Pisan were more limited (Cohen 1985).

[13] The celestial globes were calibrated on the theoretical phenomena, without taking into account the refraction, and therefore, for the verifications, I have considered the instants of rising and setting when the heights of the stars were actually 35' above the horizon (average value of refraction at the horizon).

## 8 ACKNOWLEDGEMENTS

The author would like to thank prof. Tiziana Velo for the revision of the text and Dr. Carlo Ferrigno for his invaluable work of revision and correction.

## 9 REFERENCES

Abetti G. 1949. Storia dell'astronomia. Florence: Vallecchi.
Boll F. 1901. Die Sternkataloge des Hipparch und des Ptolemaios. Bibliotheca mathematica. 3rd ser. 2:185-195.
Claudii Ptolemaei Opera quae exstant omnia. Volumen I. Syntaxis mathematica. Heiberg JL editor. 1903. Leipzig: Teubner.
Cleomedes 1891. De motu circulari corporum caelestium. Leipzig: Teubner.
Cohen IB. 1985. The birth of a new physics. New York-London: Norton & Company.
Dambis AK, Efremov YN. 2000. Dating Ptolemy's star catalogue through proper motions: the Hipparchan epoch. JHA, 31(2): 115-134.
De Lalande JJL. 1764. Astronomie. Paris: Desaint & Saillant.
Delambre JBJ. 1817. Histoire de l'astronomie ancienne. Paris: Courcier.
Diogenes Laërtius. 1853. Lives and opinions of eminent philosophers. London: Bohn.
Dobler HR. 2002. The dating of Ptolemy's star catalogue. JHA. 33(3): 265-277.
Dreyer JLE. 1917 and 1918. On the origin of Ptolemy's catalogue of stars. MNRAS. 77: 528-539 and 78 : 343-349.
Duke DW. 2002. Hipparchus' coordinate system. AHES. 56(5): 427-433.
Duke DW. 2002. Associations between the ancient star catalogues. AHES. 56(5): 435-450.
Duke DW. 2002. Dating the Almagest star catalogue using proper motion: a reconsideration. JHA. 33(1): 45-55.
Duke DW. 2003. The depth of association between the ancient star catalogues. JHA. 34(2): 227-230.
Evans J. 1987. On the origin of the Ptolemaic star catalogue. JHA. 18(3 and 4): 155-172 and 233-274.
Fedeli D. 2001. Arato, ovvero l'ipse dixit del poeta astronomo. GAst. 4: 8-13.
Graβhof G. 1990. The history of Ptolemy's star catalogue. New York: Springer Verlag.
Green DWE. 1992. Magnitude correction for atmospheric extinction. ICQ. 14: 55-59.
Gundel W. 1936. Neue astrologische Texte des Hermes Trismegistos. Munich: Verlag der Bayerischen Akademie der Wissenschaften.
Igino 2009. Mitologia astrale. Milano: Adeplhi.
Kepler J. 1610. Dissertatio cum nuncio sidereo. Prague: Sedesan.
Historia coelestis. 1672. Regensburg: Emmrich.
Kidd D. 1997. Aratus Phaenomena. Cambridge: Cambridge University Press.
Laplace PS. 1796. Exposition du sistème du monde. Paris: Imprimerie du cercle-social.
Maass E. 1892. Aratea. Philologische Untersuchungen. 12: 377-79.
Maass E. 1898. Commentariorum in Aratum reliquiae. Berlin: Weidmannos.
Manitius K. 1894. Hipparchi in Arati et Eudoxi phaenomena commentariorum libri tres. Leipzig: Teubner.

Manitius K. 1898. Gemini elementa astronomiae. Leipzig: Teubner.
Martin J. 1956. Histoire du texte des Phenomenes d'Aratos. Paris: Klincksieck.
Martin J. 1974. Scholia in Aratum Vetera. Stuttgart: Teubner.
Nadal R, Brunet J-P. 1989. Le "Commentaire" d'Hipparque II. Position de 78 étoiles. AHES. 40(4): 305-354.
Neugebauer O. 1969. The exact sciences in antiquity. New York: Dover.
Neugebauer O. 1975. A history of ancient mathematical astronomy. Berlin-Heidelberg-New York: Springer-Verlag.
Newton RR. 1977. The crime of Claudius Ptolemy. Baltimora: Johns Hopkins University Press.
Newton RR. 1979. On the fractions of degrees in an ancient star catalogue. QJRAS. 20: 383-394.
Pannekoek A. 1969. A history of astronomy (2nd ed.). New York: Barnes and Noble.
Petau D. 1630. Uranologion. Paris: Cramoisy.
Peters CHF, Knobel EB. 1915. Ptolemy's catalogue of stars: a revision of the Almagest. Washington: Carnegie Institution.
Plinio 1844. Della storia naturale. Venezia: Antonelli.
Plutarco 1991. Il volto della Luna. Milano: Adelphi.
Pseudo-Erathostenes Catasterismi. Olivieri A editor. 1897. Leipzig: Teubner.
Rawlins D. 1982. An investigation of the ancient star catalog. PASP. 94: 359-373.
Rehm A. 1899. Zu Hipparch und Eratosthenes. Hermes. 34: 251-279.
Rehm A. 1899. Eratosthenis Catasterismorum fragmenta vaticana. Ansbach: Brügel.
Russo L. 1994. The astronomy of Hipparchus and his time: a study based on preptolemaic sources. VA. 38: 207-48.
Russo L. 1996. La rivoluzione dimenticata. Milan: Feltrinelli.
Shevchenko M. 1990. An analysis of errors in the star catalogues of Ptolemy and Ulugh-Beg. JHA. 21(2): 187-201.
Strabone 1562. Della Geografia. Venice: Senese.
Swerdlow NM. 1992. The enigma of Ptolemy's catalogue of stars. JHA. 23(3): 173-183.
Toomer GJ. 1980. Hipparchus. In: Dictionary of scientific biography, vol. XV, suppl. I. New York: Scribner's sons, p. 204-227.
Toomer GJ. 1998. Ptolemy's Almagest. Princeton: Princeton University Press.
Tychonis Brahe Dani. Astronomiae instauratae progymnasmata. 1610. Prague: Brostant.
Vanin G, Cusinato B. 2017[2]. Catasterismi. L'origine, la storia, il mito delle costellazioni. Feltre: Rheticus-DBS.
Victorius P. 1567. ΙΠΠΑΡΧΟΥ ΒΙΘΥΝΟΥ ΤΩΝ ΑΡΑΤΟΥ ΚΑΥ ΕΥΔΟΞΟΥ ΘΑΙΝΟΜΕΝΩΝ ΕΞΗΓΗΣΕΩΝ βιβλία γ. Florence: Giunti.
Vitruvio 1802. Dell'architettura. Perugia: Baduel.
Vogt H. 1925. Versuch einer Wiederherstellung von Hipparchs Fixsternverzeichnis. AN. 224: 5354-5355, 17-54.
Winnecke FAT. 1878. On the visibility of stars in the Pleiades to the naked eye. MNRAS. 39: 146-148.